\documentclass[journal,twoside,web]{ieeecolor}
\usepackage{generic}
\usepackage{booktabs}
\usepackage{multirow}
\usepackage{cite}
\usepackage{amsmath,amssymb,amsfonts}
\usepackage{algorithmic}
\usepackage{graphicx}
\usepackage{textcomp}
\usepackage{scalerel}
\usepackage{tikz}
\usetikzlibrary{svg.path}
\definecolor{orcidlogocol}{HTML}{A6CE39}
\tikzset{
  orcidlogo/.pic={
    \fill[orcidlogocol] svg{M256,128c0,70.7-57.3,128-128,128C57.3,256,0,198.7,0,128C0,57.3,57.3,0,128,0C198.7,0,256,57.3,256,128z};
    \fill[white] svg{M86.3,186.2H70.9V79.1h15.4v48.4V186.2z}
                 svg{M108.9,79.1h41.6c39.6,0,57,28.3,57,53.6c0,27.5-21.5,53.6-56.8,53.6h-41.8V79.1z M124.3,172.4h24.5c34.9,0,42.9-26.5,42.9-39.7c0-21.5-13.7-39.7-43.7-39.7h-23.7V172.4z}
                 svg{M88.7,56.8c0,5.5-4.5,10.1-10.1,10.1c-5.6,0-10.1-4.6-10.1-10.1c0-5.6,4.5-10.1,10.1-10.1C84.2,46.7,88.7,51.3,88.7,56.8z};
  }
}

\newcommand\orcidicon[1]{\href{https://orcid.org/#1}{\mbox{\scalerel*{
\begin{tikzpicture}[yscale=-1,transform shape]
\pic{orcidlogo};
\end{tikzpicture}
}{|}}}}

\usepackage{hyperref}

\def\BibTeX{{\rm B\kern-.05em{\sc i\kern-.025em b}\kern-.08em
    T\kern-.1667em\lower.7ex\hbox{E}\kern-.125emX}}
\begin{document}
\title{Automated SSIM Regression for Detection and Quantification of Motion Artefacts in Brain MR Images}
\author{Alessandro Sciarra*\orcidicon{0000-0002-1247-2772}\, Soumick Chatterjee*\orcidicon{0000-0001-7594-1188}, Max D{\"u}nnwald\orcidicon{0000-0003-3838-3345}, \\ Giuseppe Placidi\orcidicon{0000-0002-4790-4029}, Andreas N{\"u}rnberger\orcidicon{0000-0003-4311-0624}, Oliver Speck\orcidicon{0000-0002-6019-5597} and Steffen Oeltze-Jafra\orcidicon{0000-0002-6962-9080}
\thanks{*Alessandro Sciarra and Soumick Chatterjee contributed equally to this work.}
\thanks{Alessandro Sciarra, Max D{\"u}nnwald and Steffen Oeltze-Jafra are with the Medicine and Digitalization - MedDigit group, Department of Neurology, Medical Faculty, University Hospital Magdeburg, Germany}
\thanks{Alessandro Sciarra, Soumick Chatterjee and Oliver Speck are with the Department of Biomedical Magnetic Resonance, Faculty of Natural Sciences, Otto von Guericke University, Germany}
\thanks{Soumick Chatterjee and Andreas  N{\"u}rnberger are with the Data and Knowledge Engineering Group, Faculty of Computer Science, Otto von Guericke University, Germany}
\thanks{Soumick Chatterjee is with the Genomics Research Centre, Human Technopole, Milan, Italy}
\thanks{Max D{\"u}nnwald is with the Faculty of Computer Science, Otto von Guericke University, Germany}
\thanks{Giuseppe Placidi is with the Department of Life, Health, and Environmental Sciences, University of L'Aquila, Italy}
\thanks{Andreas N{\"u}rnberger, Oliver Speck and Steffen Oeltze-Jafra are with the Center for Behavioral Brain Sciences, Magdeburg, Germany}
\thanks{Oliver Speck and Steffen Oeltze-Jafra are with the German Center for Neurodegenerative Disease, Magdeburg, Germany}
}

\maketitle

\begin{abstract}
Motion artefacts in magnetic resonance brain images can have a strong impact on diagnostic confidence. The assessment of MR image quality is fundamental before proceeding with the clinical diagnosis. Motion artefacts can alter the delineation of structures such as the brain, lesions or tumours and may require a repeat scan. Otherwise, an inaccurate (e.g. correct pathology but wrong severity) or incorrect diagnosis (e.g. wrong pathology) may occur. "\textit{Image quality assessment}" as a fast, automated step right after scanning can assist in deciding if the acquired images are diagnostically sufficient. 
An automated image quality assessment based on the structural similarity index (SSIM) regression through a residual neural network is proposed in this work. Additionally, a classification into different groups - by subdividing with SSIM ranges - is evaluated. Importantly, this method predicts SSIM values of an input image in the absence of a reference ground truth image. 
The networks were able to detect motion artefacts, and the best performance for the regression and classification task has always been achieved with ResNet-18 with contrast augmentation. The mean and standard deviation of residuals' distribution were $\mu=-0.0009$ and $\sigma=0.0139$, respectively. Whilst for the classification task in 3, 5 and 10 classes, the best accuracies were 97, 95 and 89\%, respectively. 
The results show that the proposed method could be a tool for supporting neuro-radiologists and radiographers in evaluating image quality quickly. 
\end{abstract}


\section{Introduction}
\label{sec:introduction}
Image quality assessment (IQA) is a fundamental tool for evaluating MR images~\cite{khosravy2019image, chow2016review, mortamet2009automatic}. The main purpose of this process is to determine if the images are diagnostically reliable and free from critical artefacts~\cite{bourel1999automatic, jezzard2009physical}. Often the evaluation process requires time and is also subjectively dependent upon the observer~\cite{ma2020diagnostic}. Furthermore, different levels of expertise and experience of the readers (experts designated to perform the IQA) could lead to variable assessment results. 
Another intrinsic issue of the IQA for MR images is the absence of a reference image. Reference-free IQA techniques with and without the machine and deep learning support have been proposed in the last years for the evaluation of the visual image quality~\cite{bourel1999automatic,chow2017modified, mortamet2009automatic, sujit2018automated, sujit2018automated, esteban2017mriqc, kustner2018automated, kustner2018machine, fantini2018automatic, fantini2021automatic,backhausen2016quality}. These techniques are able to detect and quantify the level of blurriness or corruption with different levels of accuracy and precision. However, there are many factors to take into consideration when choosing which technique to apply; the most important are~\cite{goodfellow2016deep, geron2019hands, amr2020hands}:
data requirement - as deep learning requires a large dataset while traditional machine learning (non-deep learning based) techniques can be trained on smaller data sets;
accuracy - deep learning provides higher accuracy than traditional machine learning;
training time - deep learning takes longer time than traditional machine learning;
hyperparameter tuning - deep learning can be tuned in various different ways, and it is not always possible to find the best parameters, while machine learning offers limited tuning capabilities.
In addition, when choosing traditional machine learning techniques, the fundamental step of feature extraction must be considered.
Although the list of traditional machine learning and deep learning techniques used for regression and classification tasks is constantly increasing~\cite{rawat2017deep, li2022research, staartjes2022foundations, langley2011changing}, there is still no gold standard IQA for MR images~\cite{chow2016review}.
The aim of this work is to create an automated IQA tool that is able to detect the presence of motion artefacts and quantify the level of corruption or distortion compared to an "artefact-free" counterpart based on the regression of the structural similarity index (SSIM)~\cite{wang2004image}. This tool has been designed to be able to work for a large variety of MR image contrasts, such as T1, T2, PD and FLAIR weighted images, and independently from the resolution and orientation of the considered image.
Additionally, a contrast augmentation step has been introduced in order to increase the range of variability of the weighting. In practice, when the MRIs are acquired, and there are any artefacts in the image, "artefact-free" counterparts are not available to compare the image against for quality assessment. However, for SSIM calculation, two images are required (corrupted vs motion-artefact-free images). For this reason, in this work, the corrupted images were artificially created by making use of two different algorithms - one implemented by Shaw et al.~\cite{shaw2018mri} (package of the library TorchIO~\cite{perez2021torchio}) and a second algorithm developed in-house~\cite{chatterjee2020retrospective}.
Furthermore, when training a neural network model in a fully-supervised manner, as in this case, a large amount of labelled or annotated data is typically required~\cite{atukorale2001using}. In this research on IQA, the regression labels for training were created by comparing the artificially-corrupted images against the original artefact-free images with the help of SSIM, and those SSIM values were finally used as the regression labels.


\section{Methodology}
\label{sec:methodology}
\begin{figure}
\includegraphics[width=0.5\textwidth]{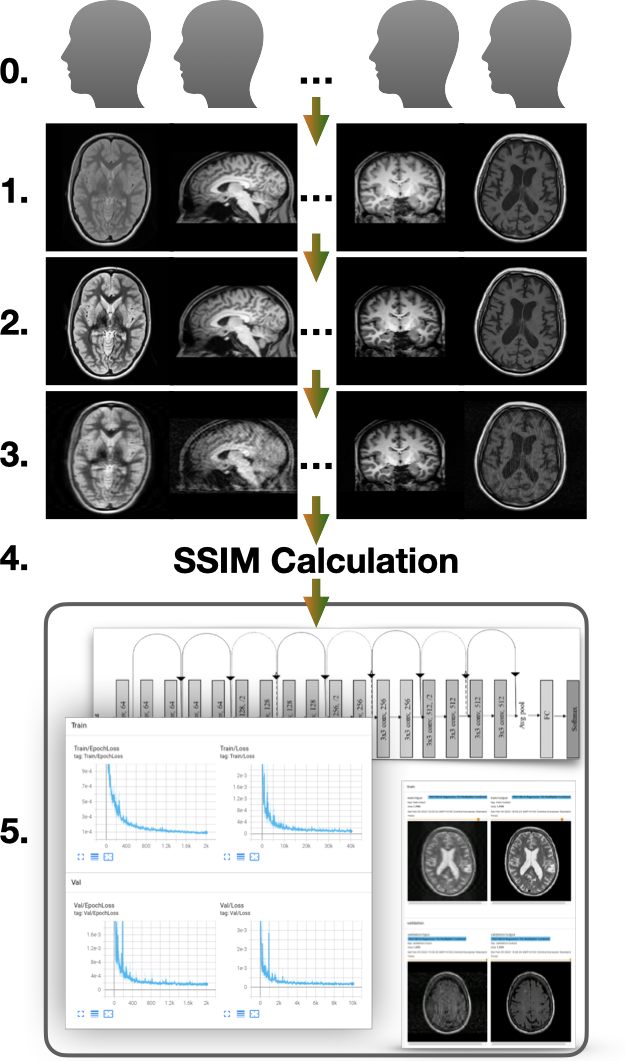}
\caption{Graphical illustration of all steps for the training as explained in Section~\ref{sec:methodology}.}
\label{PipelineFlow}
\end{figure}

\begin{figure}
\includegraphics[width=0.5\textwidth]{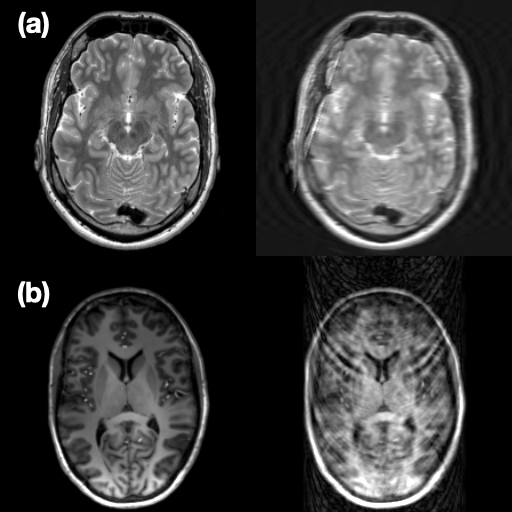}
\caption{Samples of artificially corrupted images. On the left column original images, on the right the corrupted ones. \textbf{(a)}: image corrupted making use of TorchIO library, \textbf{(b)}: image corrupted making use of the home-made algorithm }
\label{motioncorruption}
\end{figure}

The proposed automatic IQA tool relies on residual neural networks (ResNet)~\cite{7780459, NEURIPS2019_9015}. Two different versions of ResNet were used, with 18 (ResNet-18) and 101 (ResNet-101) residual blocks. Every model has been trained two times - with and without the contrast augmentation step. 
These steps are executed during the training, Figure~\ref{PipelineFlow}:
\begin{enumerate}
    \item Given a 3D input volume, one random slice (2D image) is selected from one of the possible orientations - axial, sagittal, or coronal. In the case of an anisotropic volume, the slice selection is made only following the original acquisition orientation.
    \item In case of contrast augmentation is enabled, one of the contrast augmentation algorithms is selected randomly from the following and applied to the input image:
    \begin{itemize}
        \item Random gamma adjustment~\cite{kubinger1998role}
        \item Random logarithmic adjustment~\cite{jain1995machine}
        \item Random sigmoid adjustment on the input image~\cite{braun1999image}
        \item Random adaptive histogram adjustment~\cite{pizer1987adaptive}
    \end{itemize}
    \item Motion corruption is applied on the 2D image using one of these two options:
    \begin{itemize}
        \item TorchIO~\cite{shaw2018mri, perez2021torchio}, Figure~\ref{motioncorruption} (a)
        \item "in-house" algorithm, Figure~\ref{motioncorruption} (b)
    \end{itemize}
    \item The SSIM is calculated between the 2D input image and the corresponding corrupted one.
    \item The calculated SSIM value and the corrupted image are passed to the chosen model for training
\end{enumerate}
The code for this project is available on GitHub\footnote{\url{https://github.com/sarcDV/SSIM-Class-Regression-Brain}}.

Three datasets - train, validation, and test sets - were used for this work, Table~\ref{tab0}. For training, 200 volumes were used, while 50 were used for validation and 50 for testing. The first group of 68 volumes were selected from the public IXI dataset~\footnote{Dataset available at: \underline{https://brain-development.org/ixi-dataset/}}, the second group (Table~\ref{tab0}, Site-A) of 114 volumes were acquired with a 3T scanner, the third group (Table~\ref{tab0}, Site-B) of 93 volumes was acquired at 7T,  and a final group (Table~\ref{tab0}, Site-B) of 25 volumes was acquired with different scanners (1.5 and 3T). The volumes from IXI, Site-A, and Site-B were resampled in order to have an isotropic resolution of 1.00 $mm^3$.

The loss during training was calculated using the mean squared error (MSE)~\cite{allen1971mean} and was optimised using the Adam optimiser~\cite{kingma2014adam} with a learning rate of $1e^{-3}$ and a batch size of 100 for 2000 epochs.
All images (during training, validation, and testing) were always normalised, and resized or padded to have a 2D matrix size of 256x256.

\begin{table*}
\caption{Data for training, validation and testing.}
\label{tab0}
\centering
		\begin{tabular}{c|c|c|c|c}  
        \textbf{Data} & \textbf{Weighting} &\textbf{Volumes} & \textbf{Matrix Size} & \textbf{Resolution ($mm^3$)} \\
         & & & m(M) x m(M) x m(M)\dag  & m(M) x m(M) x m(M)\dag  \\
        \midrule
        \multicolumn{5}{c}{\textbf{TRAINING}}\\
        IXI  &   T1,T2,PD      &  15,15,15 & 230(240)x230(240)x134(162) & 1.00 isotropic \\
        Site-A  &   T1,T2,PD,FLAIR      &  20,20,20,20 & 168(168)x224(224)x143(144) & 1.00 isotropic \\
        Site-B  &   T1,T2,FLAIR      &  20,20,20 & 156(156)x224(224)x100(100) & 1.00 isotropic 
        \\
        Site-C & T1 & 3 &192(512)x256(512)x36(256) & 0.45(1.00)x0.45(0.98)x0.98(4.40)\\
        Site-C & T2 & 11 &192(640)x192(640)x32(160) & 0.42(1.09)x0.42(1.09)x1.00(4.40)\\
        Site-C & FLAIR & 1 &320x320x34 & 0.72x0.72x4.40\\
        \midrule
        \multicolumn{5}{c}{\textbf{VALIDATION}}\\
        IXI  &   T1,T2,PD      &  1,5,7 & 230(240)x230(240)x134(162) & 1.00 isotropic \\
        Site-A  &   T1,T2,PD,FLAIR      &  4,4,4,4 & 168(168)x224(224)x143(144) & 1.00 isotropic \\
        Site-B  &   T1,T2,FLAIR      &  6,6,4 & 156(156)x224(224)x100(100) & 1.00 isotropic \\
        Site-C & T1 & 3 & 176(240)x240(256)x118(256) & 1.00 isotropic\\
        Site-C & T2 & 1 & 240x320x80 & 0.80x0.80x2.00\\
        Site-C & PD & 1 & 240x320x80 & 0.80x0.80x2.00\\
        \midrule
        \multicolumn{5}{c}{\textbf{TESTING}}\\
        IXI  &   T1,T2,PD      &  2,4,4 & 230(240)x230(240)x134(162) & 1.00 isotropic \\
        Site-A  &   T1,T2,PD,FLAIR      &  6,4,4,4 & 168(168)x224(224)x143(144) & 1.00 isotropic \\
        Site-B  &   T1,T2,FLAIR      &  6,6,5 & 156(156)x224(224)x100(100) & 1.00 isotropic \\
        Site-C & T1 & 2 & 288(320)x288(320)x35(46) & 0.72(0.87)x0.72(0.87)x3.00(4.40)\\
        Site-C & T2 & 2 &320(512)x320(512)x34(34) & 0.44(0.72)x0.45(0.72)x4.40(4.40)\\
        Site-C & FLAIR & 1 &320x320x35 & 0.70x0.70x4.40\\
        \bottomrule
        \multicolumn{5}{c}{\dag: "m" indicates the minimum value while "M" the maximum.}
    \end{tabular}
\end{table*}
For testing, a total of 10000 images were repetitively selected randomly and then corrupted from the 50 volumes of the test dataset - applying the random orientation selection, the contrast augmentation, and finally the corruption - as performed during the training stage.\\
In order to evaluate the performance of the trained models, foremost, the predicted SSIM values were plotted against the ground truth SSIM values as shown in Figure~\ref{regressionall}, next the residuals were calculated as follows $Residuals = SSIM_{predicted} - SSIM_{ground truth}$, Figure~\ref{regressionresiduals}.\\
The predicted SSIM value of an image can be considered equivalent to measuring the distortion or corruption level of the image. However, when applying this approach to a real clinical case, it is challenging to compare this value with a subjective assessment. To get around this problem, the regression task was simplified into a classification task. For the same, three different experiments were performed by choosing a different number of classes - 3, 5 and 10 classes. For every case, the SSIM range [0-1] was equally divided into sub-ranges. For instance, in the case of 3 classes, there were three sub-ranges, class-1:[0.00-0.33], class-2:[0.34-0.66] and class-3:[0.67-1.00]. A similar step was also performed for creating 5 and 10 classes. \\
A second dataset was also used for testing the trained models - comprised of randomly selected images from clinical acquisitions. This dataset contained five subjects, each with a different number of scans, as shown in Table~\ref{tabClinicalData}. In this case, there were no ground truth reference images, and for this reason, the images were also subjectively evaluated (subjective image quality assessment SIQA) by one expert using the following classification scheme:  class 1 - images with good to high quality that might have minor motion artefacts, but not altered structures and substructures of the brain (SSIM range between 0.85 and 1.00); class 2 - images with sufficient to good quality, in this case, the images can have motion artefacts that prevent correct delineation of the brain structures, substructures or lesions (SSIM range between 0.60 and 0.85); and class 3 - image with insufficient quality and a re-scan will be required (SSIM range between 0.00 and 0.60). Additionally, this dataset contained different contrasts not included in the training, such as diffusion-weighted images (DWI). 
As a baseline, the MRIQC\footnote{\url{https://mriqc.readthedocs.io/en/latest/about.html}}~\cite{esteban2017mriqc} toolbox has been considered for a direct comparison when applied on clinical data.
It is important to remark that MRIQC extracts several no-reference image quality metrics only from T1w and T2w 3D image volumes and functional MRI data. 
For this reason, many of the clinical volumes were discarded during the quality assessment.
Furthermore, MRIQC could not be used for the assessment of the artificially corrupted images because it works only on acquisitions properly converted to the BIDS \footnote{\url{https://bids-specification.readthedocs.io/en/stable/index.html}} format (i.e. artificially corrupted 2D slices are not suitable for MRIQC). The following metrics for structural images were used:
the contrast-to-noise ratio (CNR)~\cite{magnotta2006measurement}, the coefficient of joint variation (CJV)~\cite{ganzetti2016intensity}, the entropy focus criterion (EFC)~\cite{atkinson1997automatic} and the so called quality index (QI)~\cite{mortamet2009automatic}.
The first one, CNR, is a common image metric and a simple extension of the signal-to-noise ratio (SNR) calculation; it is able to assess the separation between the tissue distributions of grey and white matter (GM and WM). Higher values indicate better image quality.
The second chosen metric, CJV, can detect the presence of heavy head motion and large intensity non-uniformities (INU), and for this metric lower values indicate better image quality.
One of the earliest proposed metrics available in MRIQC is the EFC. With this metric is possible to quantify the level of ghosting and blurriness caused by the head motion. It makes use of the Shannon entropy of voxel intensities. Images with lower EFC values present a better image quality.
The last quality measure, QI, is a binary metric that reflects the presence or absence of artefacts.
When QI is non-zero, the image is corrupted by artefacts, while zero QI indicates an absence of artefacts. Those metrics were chosen among the others for their specificity towards artefact detection and quantification.
In order to assess the agreement with SIQA results, for each chosen metric, the SIQA scores were specifically averaged, normalised and scaled. 
The averaging step is required since the SIQA scores were per slice, while MRIQC reports a single number for each metric of every scan. 
For the first three metrics, CNR, CJV and EFC, the SIQA scores were normalised and scaled, taking as reference the first analysed image volume by MRIQC.
Whilst for the QI metric, the averaged SIQA scores within the range $1-2$ were converted to zero values to indicate the absence of motion artefacts as for the QI metric; otherwise, 1 to report the presence of artefacts. 
\begin{table*}
\caption{Clinical data}
\label{tabClinicalData}
\centering
	\begin{tabular}{c|c|c|c|c}
        \textbf{Data} & \textbf{Weighting} &\textbf{Volumes} & \textbf{Matrix Size} & \textbf{Resolution ($mm^3$)}  \\
         & & & m(M) x m(M) x m(M)\dag & m(M) x m(M) x m(M)\dag\\
        \midrule
        Subj. 1  & T1,T2,FLAIR     &  1,4,2 & 130(560)x256(560)x26(256) & 0.42(1.00)x0.42(0.94)x0.93(4.40) \\
        Subj. 2  & T2 & 3 & 288(320)x288(320)x28(28) & 0.76(0.81)x0.76(0.81)x5.50(5.50)  \\
        Subj. 3  & T1,T2,FLAIR,DWI,(\S) &  1,2,1,4,1 &  256(640)x256(640)x32(150) & 0.42(0.90)x0.42(0.90)x0.45(4.40) \\
        Subj. 4  & T2, FLAIR, DWI & 1,2,6 & 144(512)x144(512)x20(34) & 0.45(1.40)x0.45(1.40)x2.00(4.40)\\
        Subj. 5 & T2, FLAIR, DWI & 3,1,4 & 256(640)x256(640)x28(42) & 0.40(1.09)x0.40(1.09)x3.30(6.20)  \\
        \bottomrule
        \multicolumn{5}{c}{\dag: "m" indicates the minimum value while "M" the maximum.}
    \end{tabular}
\end{table*}
\section{Results}
\label{sec:results}
\begin{figure}
\centering
\includegraphics[width=0.5\textwidth]{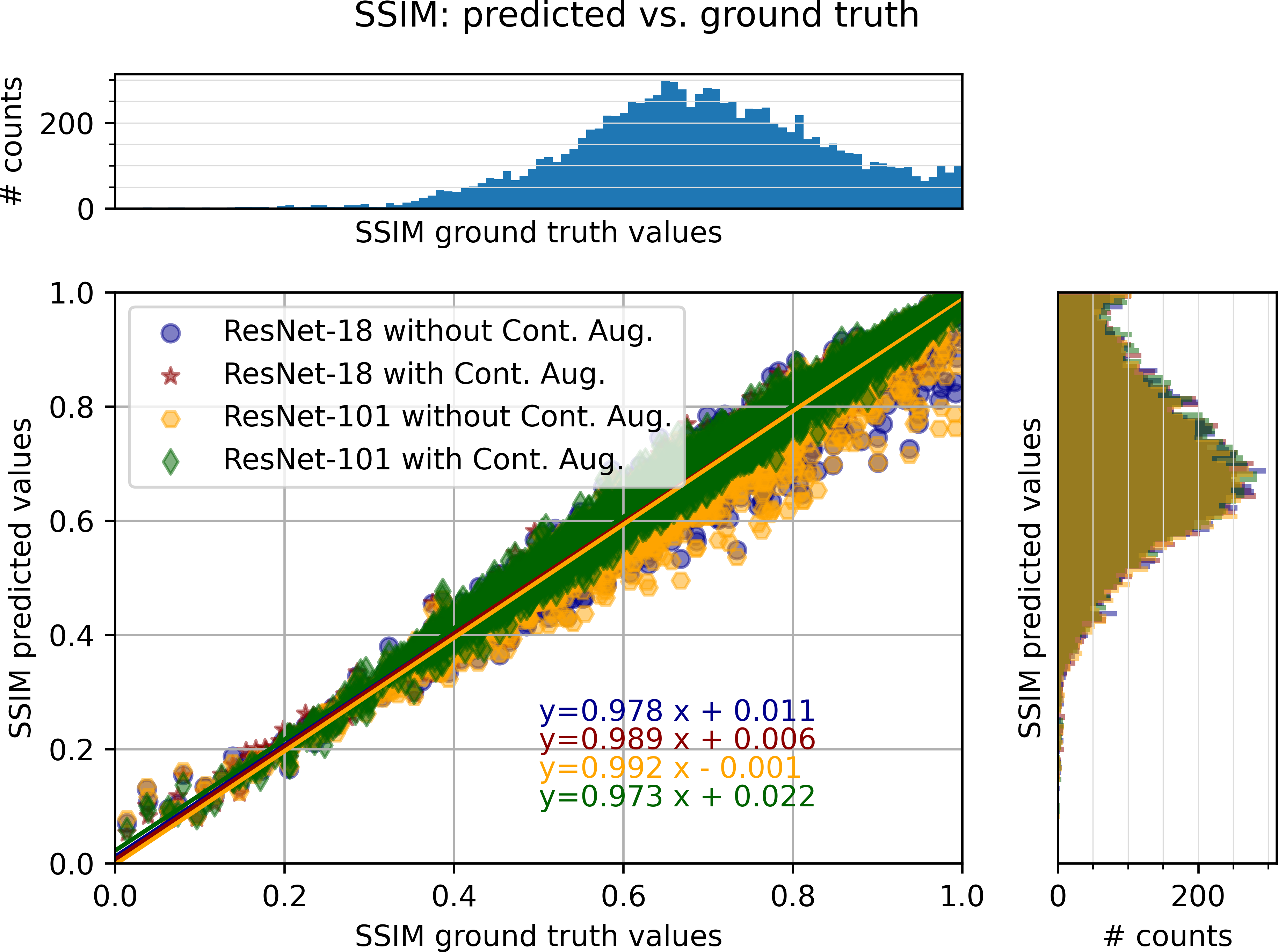}
\caption{Scatter plot for the regression task. There are also shown the linear fittings for each group of data. On the top: ground truth SSIM values distribution; right side: predicted SSIM values distributions for each group of data.}
\label{regressionall}
\end{figure}

\begin{figure*}
\includegraphics[width=\textwidth]{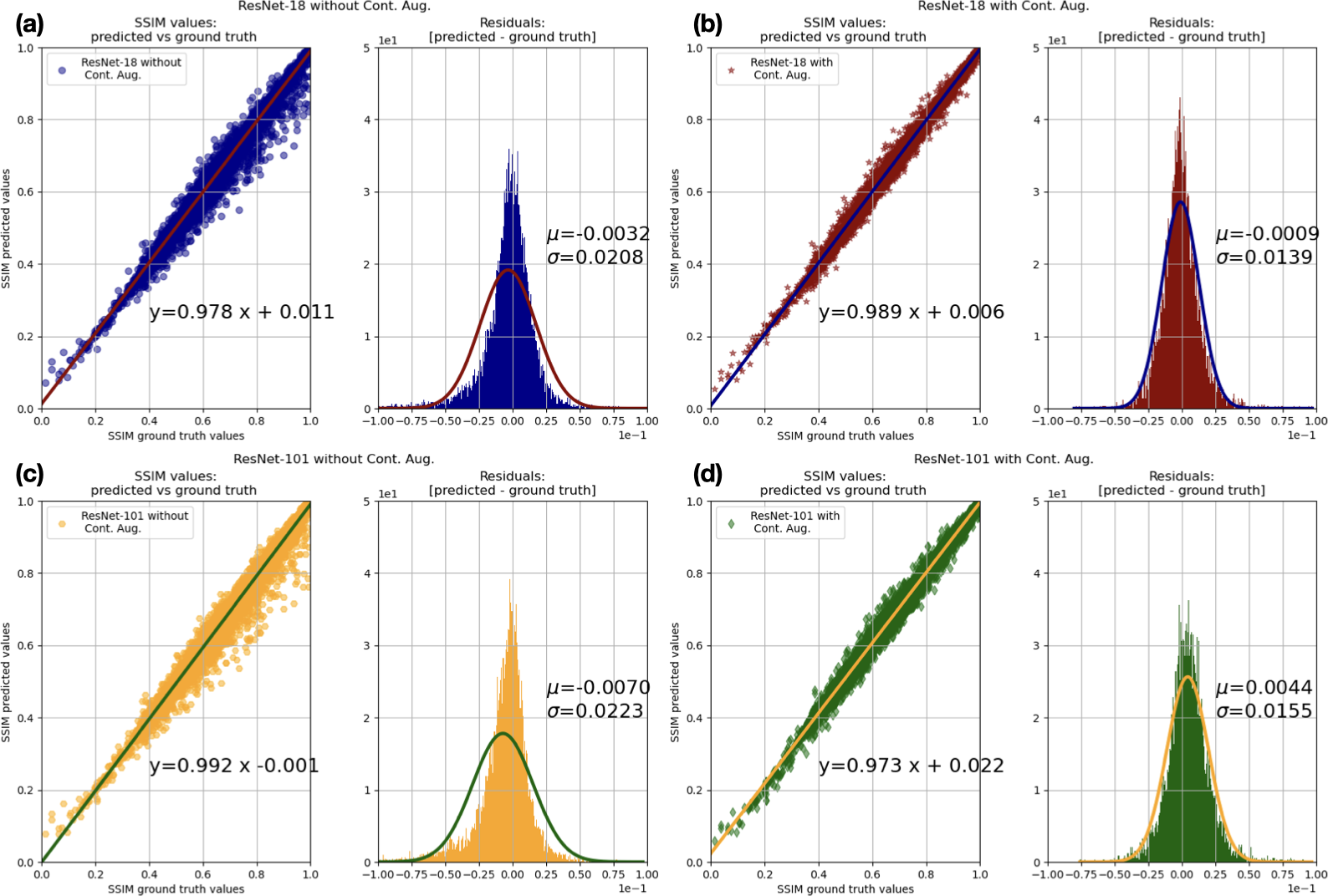}
\caption{Scatter plot SSIM predicted against ground truth values and Residuals distribution for (a) ResNet-18 without contrast augmentation, (b) ResNet-18 with contrast augmentation, (c) ResNet-101 without contrast augmentation and (d) ResNet-101 with contrast augmentation. }
\label{regressionresiduals}
\end{figure*}

\begin{figure*}
\includegraphics[width=\textwidth]{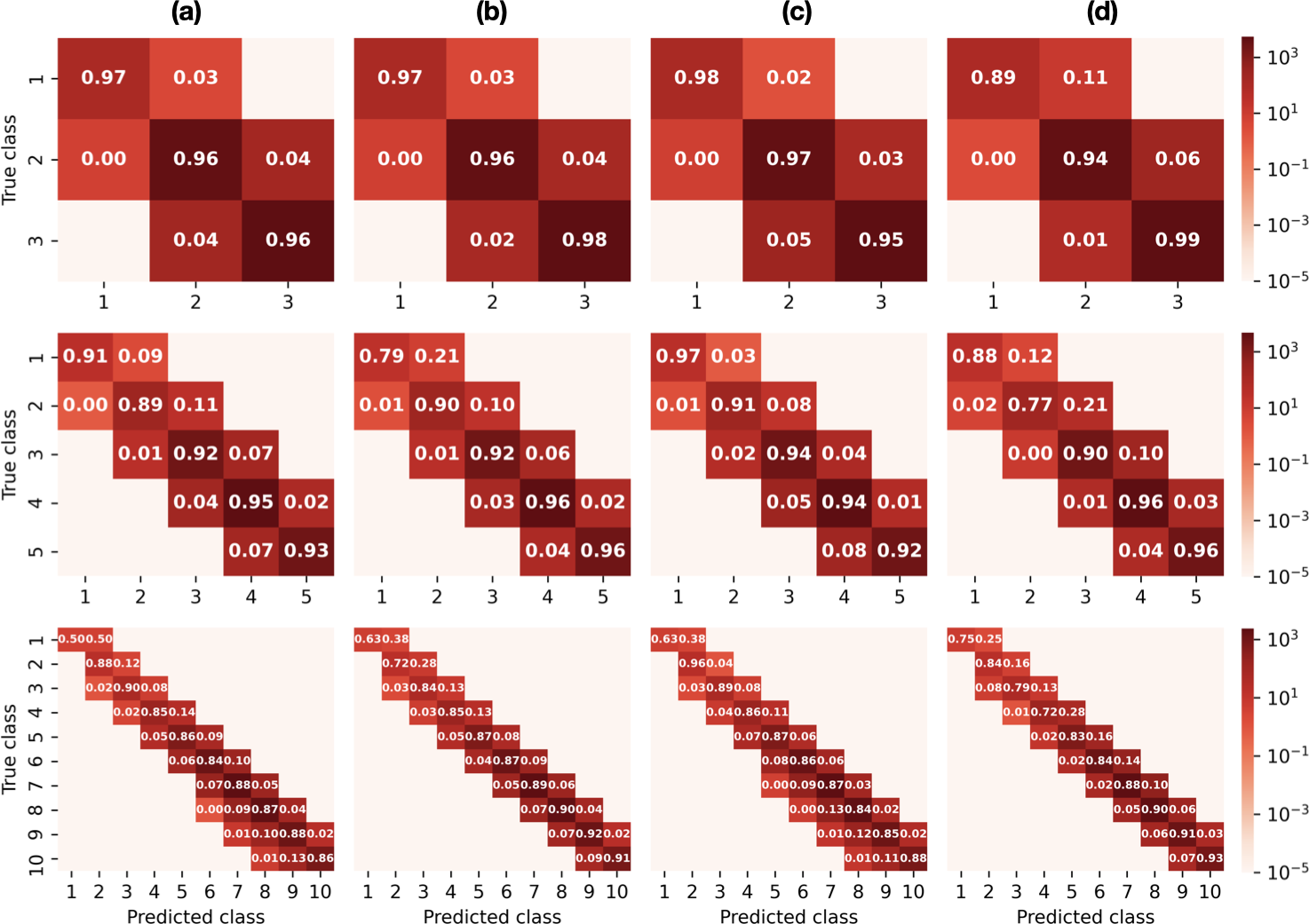}
\caption{Confusion matrices for the classification task. First row 3-class case, second row 5-class and third row 10-class scenario. The columns are for \textbf{(a)} ResNet-18 without contrast augmentation, \textbf{(b)} ResNet-18 with contrast augmentation, \textbf{(c)} ResNet-101 without contrast augmentation, \textbf{(d)} ResNet-101 with contrast augmentation, respectively.}
\label{threeClassification}
\end{figure*}

The results for the first section, the regression task, are presented in Figures~\ref{regressionall} and~\ref{regressionresiduals}. Figure~\ref{regressionall} shows a scatter plot where the predicted SSIM values are compared against the ground truth values. Additionally, the plot shows the linear fitting performed for each trained model. Finally, the distributions of the ground truth and predicted SSIM values are also shown.
Figure~\ref{regressionall} presents general comparisons across all the trained models and their qualitative dispersion levels. In this case, the term dispersion implies how much the predicted SSIM values differ from the ground-truth $SSIM_{predicted}= SSIM_{ground-truth}$. On the other hand, in Figure~\ref{regressionresiduals}, the results are shown separately using the scatter plots - for each model. The relative residual distribution plots are explained in section~\ref{sec:methodology}. 
For the residual distributions, a further statistical normal distribution fitting was carried out, making use of the python package SciPy~\cite{2020SciPy-NMeth}. The calculated mean and standard deviation values are shown in Figure~\ref{regressionresiduals}.
According to the statistical analysis, the model that has the smallest standard deviation ($\sigma = 0.0139$) and the mean value closest to zero ($\mu = -0.0009$) was the ResNet-18 model trained with contrast augmentation, while the model with the mean value furthest from zero and largest standard deviation was the  ResNet-101 trained without contrast augmentation.
A clear effect of the contrast augmentation for both models, ResNet-18 and ResNet-101, can be seen in the results - reflected as a reduction of the standard deviation values, and this visually correlates with a lower dispersion level in the scatter plots. 
The results for the classification task are shown in Figure~\ref{threeClassification} and table~\ref{tab1}. 
Figure~\ref{threeClassification} shows the logarithmic confusion matrices obtained for the classification task. It can be noted that all the trained models performed well and in a similar way. In particular, none of the matrices presents non-zero elements far from the diagonal, but only the neighbouring ones - as commonly expected from a classification task. 
The table~\ref{tab1} is complementary to Figure~\ref{threeClassification}. It shows the class-wise, macro-average and weighted average of precision, recall, and f1-score for all the trained models. This table also presents the accuracy. For all the three scenarios, 3, 5 and 10 classes as presented in section~\ref{sec:methodology}, once again, the model with the best results is ResNet-18 trained with contrast augmentation. This model always obtained the highest accuracy value - 97, 95 and 89\% for 3, 5, and 10 class scenarios, respectively.
Even though the ResNet-18 with contrast augmentation performed better than the other models, no substantial differences can be discerned from the tabular data. But once again, it is possible to observe an improvement in terms of performance when contrast augmentation is applied.
%
\begin{table*}
\caption{Results for the classification task. The classification task has been performed three times, considering 3,5 and 10 classes, respectively. "Prec." is the abbreviation of the term precision, while "macro avg" corresponds to macro average and "weight. avg" to the weighted average calculated using the python package scikit-learn~\cite{scikit-learn}. \textbf{(a)} is for ResNet-18 without contrast augmentation,  \textbf{(a)} is for ResNet-18 with contrast augmentation,\textbf{(c)} is for ResNet-101 without contrast augmentation, \textbf{(c)} is for ResNet-101 with contrast augmentation.}
\label{tab1}
\centering
	\begin{tabular}{l | ccc |  ccc | ccc | ccc |c}
		    & & \textbf{(a)} & & & \textbf{(b)} & & & \textbf{(c)}& & & \textbf{(d)} & &  \\
		\midrule
		Class (SSIM)& Prec. & Recall & f1-score & Prec. & Recall & f1-score & Prec. & Recall & f1-score &  Prec. & Recall & f1-score & Support \\
		\midrule
		1 [0.00 - 0.33] & 0.94 & 0.97 & 0.95 & 0.93 & 0.97 & 0.95 & 0.93 & 0.98 & 0.96 & 0.97 & 0.89 & 0.93 & 117 \\
		2 [0.033 - 0.66]& 0.95 & 0.96 & 0.95 & 0.97 & 0.96 & 0.96 & 0.94 & 0.97 &  0.95 & 0.98 & 0.94 & 0.96 & 4307 \\
		3 [0.66 - 1.00]& 0.97 & 0.96 & 0.97 & 0.97 & 0.98 & 0.97 & 0.98 & 0.95 & 0.96 & 0.95 & 0.99 & 0.97 & 5576 \\
		    & & & & & & & & & & & & &  \\
		accuracy &  &  & 0.96 &  &  & \textbf{0.97} & & & 0.96 & & & 0.96 & 10000 \\
		macro avg &  0.95 & 0.95 & 0.96 & 0.96 & 0.97 & 0.96 & 0.95 & 0.97 & 0.96 & 0.97 & 0.94 & 0.95 & 10000 \\
		weight. avg & 0.96 & 0.96  & 0.96 & 0.97 & 0.97 & 0.97 & 0.96 & 0.96 & 0.96 & 0.96 & 0.96 & 0.96 & 10000 \\
		\midrule
		1 [0.00 - 0.20] & 0.97 & 0.91 & 0.94 & 0.93 & 0.79 & 0.85 & 0.94 & 0.97 & 0.96 & 0.85 & 0.88 & 0.87 &  33 \\
		2 [0.20 - 0.40]& 0.86 & 0.89 & 0.88 & 0.85 & 0.90 & 0.87 & 0.83 & 0.91 & 0.87 & 0.93 & 0.77 & 0.84 & 262 \\
	    3 [0.40 - 0.60]& 0.91 & 0.92 & 0.91 & 0.93 & 0.92 & 0.93 & 0.89 & 0.94 & 0.91 & 0.94 & 0.90 & 0.92 & 2320 \\
		4 [0.60 - 0.80]& 0.94 & 0.95 & 0.94 & 0.95 & 0.96 & 0.96 & 0.94 & 0.94 & 0.94 & 0.94 & 0.96 & 0.95 & 5021 \\
		5 [0.80 - 1.00]& 0.96 & 0.93 & 0.95 & 0.96 & 0.96 & 0.96 & 0.97 & 0.92 & 0.95 & 0.95 & 0.96 & 0.96 & 2364 \\
		    & & & & & & & & & & & & &  \\
		accuracy & & & 0.93 & & & \textbf{0.95} & & & 0.93 & & & 0.94 & 10000 \\
		macro avg & 0.93 & 0.92 & 0.92 & 0.93 & 0.91 & 0.91 & 0.91 & 0.93 & 0.92 & 0.92 & 0.89 & 0.91 & 10000 \\
		weight. avg & 0.93 & 0.93 & 0.93 & 0.95 & 0.95 & 0.95 & 0.93 & 0.93 & 0.93 & 0.94 & 0.94 & 0.94 & 10000 \\
		\midrule
		1 [0.00 - 0.10] & 1.00 & 0.50 & 0.67 & 1.00 & 0.62 & 0.77 & 1.00 & 0.62 & 0.77 & 1.00 & 0.75 & 0.86 & 8 \\
		2 [0.10 - 0.20]& 0.81 & 0.88 & 0.85 & 0.78 & 0.72 & 0.75 & 0.83 & 0.96 & 0.89 & 0.75 & 0.84 & 0.79 & 25 \\
		3 [0.20 - 0.30]& 0.90 & 0.90 & 0.90 & 0.81 & 0.84 & 0.83 & 0.87 & 0.89 & 0.88 & 0.91 & 0.79 & 0.84 & 62 \\
		4 [0.30 - 0.40]& 0.81 & 0.84 & 0.83 & 0.80 & 0.85 & 0.83 & 0.76 & 0.85 & 0.80 & 0.88 & 0.71 & 0.79 & 200 \\
		5 [0.40 - 0.50]& 0.82 & 0.86 & 0.84 & 0.86 & 0.87 & 0.87 & 0.79 & 0.87 & 0.83 & 0.86 & 0.83 & 0.84 & 689 \\
		6 [0.50 - 0.60]& 0.84 & 0.84 & 0.84 & 0.89 & 0.87 & 0.88 & 0.83 & 0.86 & 0.84 & 0.89 & 0.84 & 0.86 & 1631 \\
		7 [0.60 - 0.70]& 0.86 & 0.88 & 0.87 & 0.89 & 0.89 & 0.89 & 0.85 & 0.87 & 0.86 & 0.88 & 0.88 & 0.88 & 2706 \\
		8 [0.70 - 0.80]& 0.87 & 0.87 & 0.87 & 0.89 & 0.90 & 0.89 & 0.88 & 0.84 & 0.86 & 0.86 & 0.90 & 0.88 & 2315 \\
		9 [0.80 - 0.90]& 0.86 & 0.88 & 0.87 & 0.89 & 0.92 & 0.90 & 0.89 & 0.85 & 0.87 & 0.87 & 0.91 & 0.89 & 1456 \\
		10 [0.80 - 1.0]& 0.97 & 0.86 & 0.91 & 0.97 & 0.91 & 0.94 & 0.96 & 0.88 & 0.91 & 0.95 & 0.93 & 0.94 & 908 \\
		    & & & & & & & & & & & & &  \\
		accuracy & & & 0.87 & & & \textbf{0.89} & & & 0.86 & & & 0.88 & 10000 \\
		macro avg & 0.88 & 0.83 & 0.84 & 0.88 & 0.84 & 0.85 & 0.86 & 0.85 & 0.85 & 0.88 & 0.84 & 0.86 & 10000 \\
		weight. avg & 0.87 & 0.87 & 0.87 & 0.89 & 0.89 & 0.89 & 0.86 & 0.86 & 0.86 & 0.88 & 0.88 & 0.88 & 10000 \\
		\bottomrule
		\end{tabular}
\end{table*}
\begin{figure*}
\centering
\includegraphics[width=\textwidth]{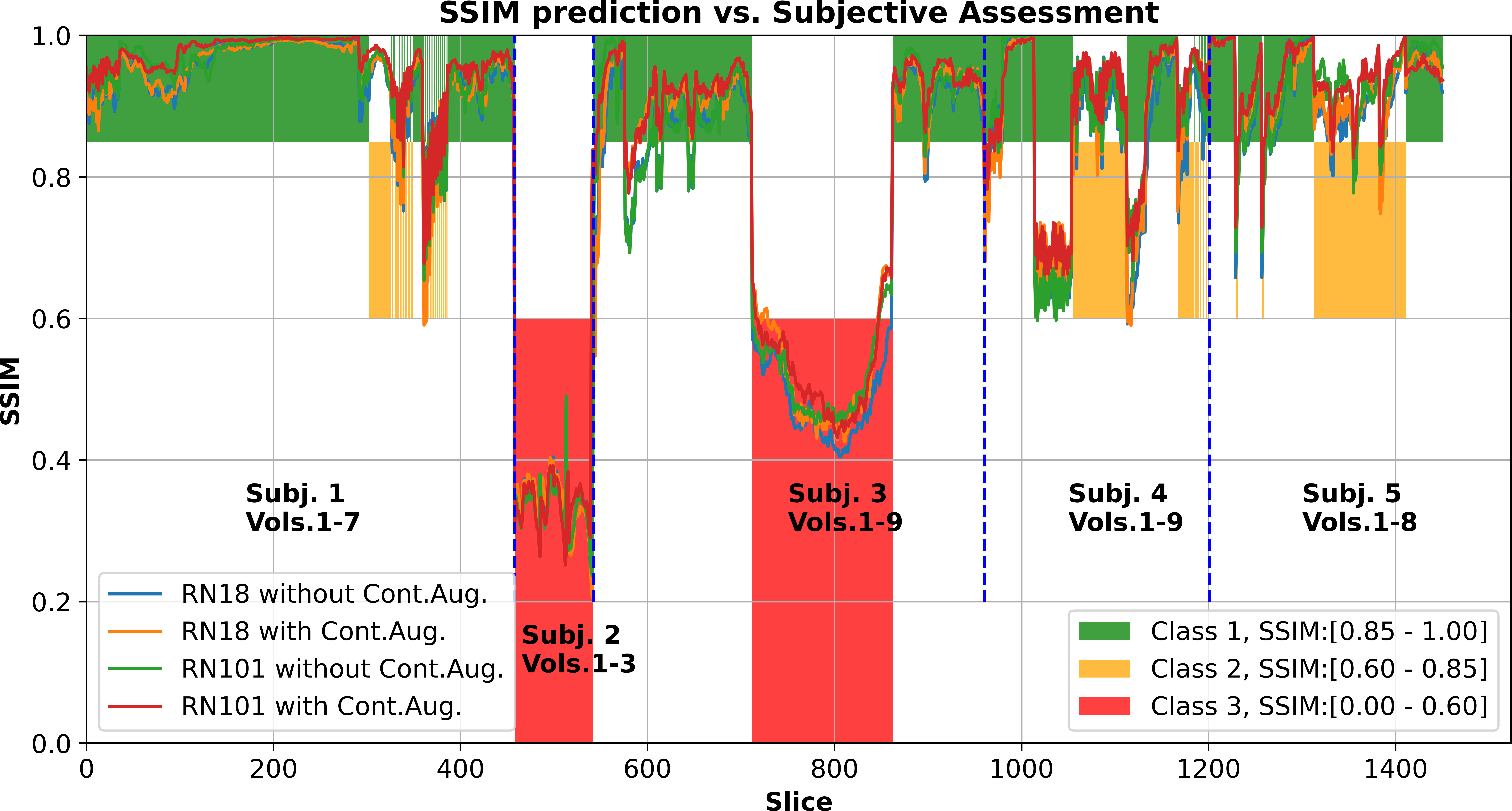}
\caption{Evaluation for the clinical dataset. The curves represent the SSIM predictions obtained with the different trained models, while the coloured bars show the subjective classification performed by the expert. When the curves are within the coloured bars, there is an agreement between the objective and subjective evaluation, disagreement otherwise. The blue dashed lines indicate the separation between the different subjects.
On the x-axis, there is the slice number; all the volumes were stacked consecutively one after another.}
\label{clinicalResults}
\end{figure*} 
The results regarding the clinical data samples are shown in Figure~\ref{clinicalResults}. In this case, the obtained SSIM predictions are shown for each model - overlayed with the subjective scores - shown in a per-slice manner grouped by the subjects.
As introduced in section~\ref{sec:methodology}, the subjective ratings for the clinical data samples were within the classes 1, 2 or 3 - after a careful visual evaluation.
If the predictions obtained with the different models fall within the classes assigned by the subjective evaluation, this implies that there is an agreement between the objective and subjective evaluations. When the objective prediction lies outside the class assigned by the expert, this indicates a disagreement between the two assessments.
The percentage of agreement between subjective and objective analysis is $76.6\pm0.8\%$ (mean $\pm$ standard deviation), with a minimum value of $75.5\%$ achieved by ResNet-101 without contrast augmentation and a maximum of $77.7\%$ by ResNet-101 with contrast augmentation.
\begin{figure*}
	\centering
	\includegraphics[width=\textwidth]{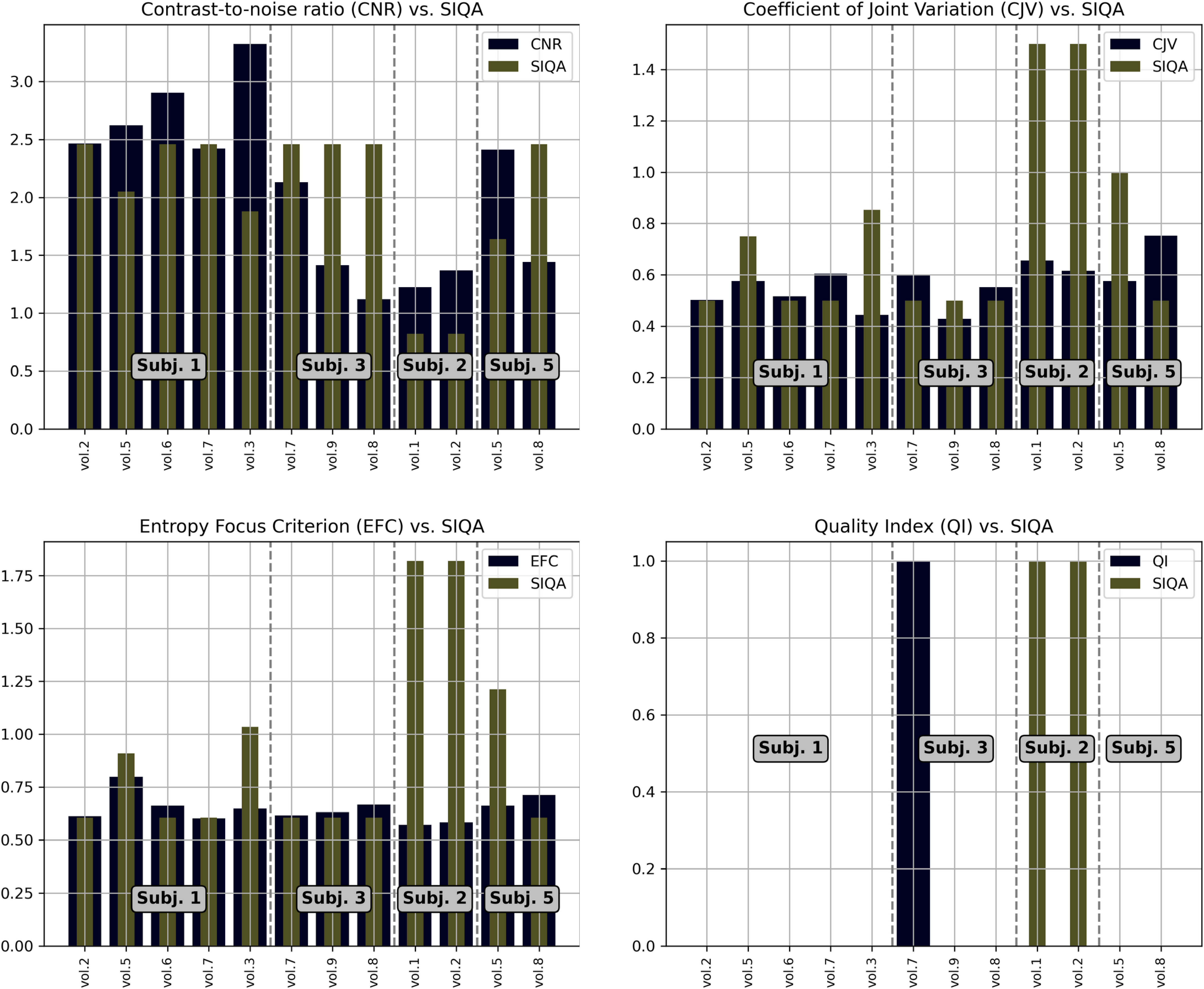}
	\caption{MRIQC results. Top left: CNR; top right: CJV; bottom left: EFC, and bottom right: QI. }
	\label{MRIQCResults}
\end{figure*}
The results from MRIQC are presented in figure~\ref{MRIQCResults}. Once again, it is important to mention that MRIQC is a toolbox which comprises several image quality metrics and provides a comprehensive analysis of the scans. However, only 12 of 36 total scans were processed, mainly because the clinical scans were not T1w or T2w acquisitions as required by MRIQC.
The rates of agreement between the chosen MRIQC metrics and the SIQA scores were the following:
17\%, 17\%, 33\% and 75\% for CNR, CJV, EFC and QI, respectively.
\section{Discussion}
\label{sec:discussion}
The performances of the trained models when solving the regression task were very similar. However, both models ResNet-18 and ResNet-101, showed a distinct improvement when coupled with contrast augmentation.
Looking at the Residuals distributions of the errors for both models, contrast augmentation has been the reason why the mean values fell closer to zero and also, the values of the standard deviation decreased by $\approx 1.5$ and $\approx 1.44$ times for ResNet-18 and ResNet-101, respectively. 
The reduction of the standard deviations is also evident in the scatter plots, where the dispersion level is visibly less when contrast augmentation is applied.
While considering the classification task, the first notable thing is that there is a linear decrease in the accuracy as the number of classes increases - $97$, $95$ and $89\%$. This can be explained by the fact that as the number of classes increases, the difficulty level also increases for each model to classify the image in the correct pre-defined range of SSIM values. The confusion matrices confirm this behaviour - by the increase of the values being out-of-diagonal, i.e., considering the ResNet-18 not coupled with contrast augmentation, for the classification task with three classes, the maximum value out-of-diagonal is 0.04 (for class-2 and class-3), while considering the classification task with ten classes, the maximum value is 0.50 (for class-1). This implies that the ResNet-18, not coupled with contrast augmentation when performing the 10-classes classification task, classifies incorrectly $50\%$ of the tested images. When contrast augmentation is applied, there is an apparent reduction of wrongly classified images of class-1. Although this is the general trend observed in Figure~\ref{threeClassification}, there are also contradicting results, i.e., when looking at the 5-classes classification task for class-1 always considering ResNet-18 without and with contrast augmentation, there is a net increase of erroneously classified class-1 images, from $9$ to $21\%$ of tested images.
The final application on clinical data also provided satisfactory results, with a maximum agreement rate of $77.7\%$ between the objective and subjective assessments.
A direct comparison with the previous three-classes classification task is not possible due to the different subjective schemes selected (Section~\ref{sec:methodology}).
Although there is a visible reduction in performance when the trained models are applied to clinical data, this can be justified by taking into account several factors. First of all, the clinical data sample involved types of image data, such as diffusion acquisition and derived diffusion maps, which were never seen by the models during the training step, and secondly, the motion artefacts artificially created did not cover the infinite possible motion artefacts that can appear in a truly MR motion corrupted image. 
A possible improvement can be obtained by introducing new contrasts in the training set, different resolutions and orientations. For example, oblique acquisitions have not been considered in this work. 
In addition, the artificial corruption methods used for this work can be further improved, e.g., including corruption algorithms based on motion log information recorded by a tracking device, as commonly used for prospective motion correction~\cite{herbst2014reproduction},~\cite{zahneisen2016reverse},~\cite{sciarra2022quantitative}. However, this would require the availability of raw MR data, and it has to be taken into account also the computational time to de-correct the images, comparably slower than the current approaches.
Another point to take into account for the subjective assessment is the bias introduced by each expert while evaluating the image quality. 
In this work, the expert’s perception of image quality is emulated with good accuracy, $76.6\pm0.8\%$, which can not be considered a standard reference.
Although the subjective assessment can be repeated with the help of several experts, there will always be differences between them, i.e., years of experience or different sensitivity to the presence of motion artefacts in the assessed image.
It is also noteworthy that the SSIM ranges defined for the three classes can be re-defined following a different scheme. 
In the scenario explored in this paper, the scheme has been defined by making use of the artificially corrupted images and the ground truth images - this allowed an exact calculation of the SSIM values, and it was simple to define ranges that visually agree with the scheme defined in Sect.~\ref{sec:methodology}.
%
The results of MRIQC seem to be less in agreement with the SIQA, at least for three metrics, CNR, CJV and EFC. However, the QI measure has a rate agreement of 75\% with SIQA and taking into consideration only the scans analysed by MRIQC, the rate agreement between the QI measure and our method is also 75\%.

\section{Conclusion}
\label{sec:conclusion}
This research presents an SSIM-regression-based IQA technique using ResNet models, coupled with contrast augmentations, to make them robust against changes in image contrasts in clinical scenarios. 
The method managed to predict the SSIM values from artificially motion-corrupted images without the ground-truth (motion-free) images with high accuracy (residual SSIMs as less as $-0.0009\pm0.0139$). 
Moreover, the motion classes obtained from the predicted SSIMs were very close to the true ones and achieved a maximum weighted accuracy of $89\%$ for the ten classes scenario as reported in Table~\ref{tab1}, and achieved a maximum accuracy value of $97\%$ when the number of classes was three (Table~\ref{tab1}). 
Considering the complexity of the problem in quantifying the image degradation level due to motion artefacts and additionally the variability of the type of contrast, resolution, etc., the results obtained are very promising.
Further evaluations, including multiple subjective evaluations, will be performed on clinical data to judge its clinical applicability and robustness against changes in real-world scenarios.
In addition, other training will be carried out in order to have a larger variety of images that should include common clinical routine acquisitions such as diffusion-weighted imaging and Time-of-Flight imaging. Furthermore, it would be beneficial to include images also acquired at lower magnetic field strength ($< 1.5$ T). Considering the results obtained by ResNet models in this work, it is reasonable to think that future works can also be targeted towards a different anatomical body part, focusing, for instance, on abdominal or cardiac imaging. 
However, the reproduction of real-looking-like motion artefacts plays a key role in the performances of deep learning models trained to have a reference-less image quality assessment tool.
%
\begin{table*}
\caption{Comparison table: MRIQC (baseline) and our A-SSIM-Regr.
        $^a$Hardware required for clinical data evaluation.
        $^{\dagger}$ optional but highly recommended for training a new model.
        $^{*}$ Docker size. 
        $^{**}$ MRIQC could not process all the clinical image volumes, only the structural ones, T1w and T2w.}
\label{tab2}
\centering
	\begin{tabular}{l | c | c}
	                & MRIQC & Ours \\ 
\midrule
Data preparation  & Mandatory BIDs   & Not required, any format can be used: \\
                  & conversion       & DICOM, Nifti, etc. \\
\midrule
RAM/ROM$^a$ required & 49 GB / $\approx$ 16 GB$^{*}$ & 4GB /  \\
VRAM (on GPU) &  - & $\approx$ 1 GB \\
CPU$^a$  &  AMD Ryzen 9 (boost up to 4.7GHz)  &  Intel® Core™ i7-8700K \\
GPU$^a$  &  Not required & NVIDIA GeForce GTX 1080 Ti$^{\dagger}$\\
\midrule 
Time required (CPU) &  15 minutes to assess 12 vol.$^{**}$  & 39.79 seconds for 36 vol. \\
Time required (GPU) &  Not available  &  8.84 seconds for 36 vol.\\
\midrule
Type of images & only 3D T1w, T2w and fMRI & ALL (2D and 3D)\\
\midrule
Dependencies & FSL, ANTs, AFNI, FreeSurfer, etc. & Python, PyTorch \\
             & Docker alternative is available & \\
\midrule
Image Quality Metrics  & Multiple (CNR, CJV, EFC, etc.) & Single \\
\bottomrule
	\end{tabular}
\end{table*}
%
\bibliographystyle{IEEEtran}
\bibliography{bibfile}
\end{document}